\documentclass[twocolumn,amsmath,amssymb,letter]{revtex4-1}

\usepackage{bm}
\usepackage[dvipdfmx]{graphicx}
\usepackage{color}
\usepackage{braket}

\usepackage{mathdots}
\usepackage[pdftex]{hyperref}

\hypersetup{
	colorlinks=true
}

\newif\iftitle
\titlefalse

\newcommand  {\eqn}[1]{(\ref{eqn:#1})}
\renewcommand{\(}     {\left(}
\renewcommand{\)}     {\right)}
\renewcommand{\[}     {\left[}
\renewcommand{\]}     {\right]}
\renewcommand{\_}[1]  {_\textrm{#1}}

\begin{document}

\title{
Ballistic transport in disordered Dirac and Weyl semimetals
}

\author{Koji Kobayashi$^1$}
\author{Miku Wada$^2$}
\author{Tomi Ohtsuki$^2$}
\affiliation{$^1$Institute for Materials Research, Tohoku University, Sendai Aoba-ku 980-8577, Japan}
\affiliation{$^2$Physics Division, Sophia University, Chiyoda-ku, Tokyo 102-8554, Japan}



\begin{abstract}
 We study the dynamics of Dirac and Weyl electrons in disordered point-node semimetals.
 The ballistic feature of the transport is demonstrated
by simulating the wave-packet dynamics on lattice models.
 We show that the ballistic transport survives under a considerable strength of disorder up to the semimetal-metal transition point,
which indicates the robustness of point-node semimetals against disorder.
 We also visualize the robustness of the nodal points and linear dispersion under broken translational symmetry.
 The speed of the wave packets slows down with increasing disorder strength,
and vanishes toward the critical strength of disorder,
hence becoming the order parameter.
 The obtained critical behavior of the speed of the wave packets is consistent with that predicted by the scaling conjecture.
\end{abstract}

\maketitle

\iftitle
\section{Introduction} \label{sec:intro}
\else
 \textit{Introduction.}
\fi
 Dirac/Weyl semimetals (DSM/WSM) \cite{Murakami07phase,Burkov11weyl,Wan11topological,Okugawa14dispersion} are three-dimensional (3D) systems where an electron near the Fermi energy obeys the massless Dirac/Weyl-like equation of motion.
 They show 3D linear dispersions and nodal points,
and are called point-node semimetals (PNSMs).
 The PNSMs are often defined by a vanishing density of states (DOS) at the nodal point.
 However, in the presence of disorder, the problem of DOS is complicated because ``rare events'' might introduce a small but finite DOS \cite{Nandkishore14rare,Syzranov15unconventional,Pixley16rare,Sbierski17quantitative,Buchhold18vanishing,Buchhold18nodal}.
 It is still an open question whether or not the DOS at the nodal point is finite;
in other words, we have not reached a clear agreement whether or not ``disordered PNSM'' exists
\cite{Goswami11quantum,Sbierski14quantum,Skinner14coulomb,Roy14diffusive,Roy16erratum,Syzranov15criticalTransport,Pixley15anderson,Sbierski15quantum,Chen15disorder,Liu16effect,Bera16dirty,Pixley16disorder,Shapourian16phase,Pixley16uncovering,Louvet16on,Ziegler16quantum,Louvet17new,Wilson17quantum,Pixley17single,Syzranov18high,Wilson18do,Klier19from,Brillaux19multifractality,Syzranov19duality,Wilson20avoided}.
 The problem seems difficult to resolve, since the investigation of the vanishing DOS exactly at the nodal point is 
difficult both theoretically and experimentally.
 Therefore, it would be desirable to characterize the disordered PNSMs 
by other observables.

 In this paper, we \textit{visualize} a characteristic feature of disordered PNSMs:
ballistic transport under disorder.
 The simulation of wave-packet dynamics clearly shows the ballistic feature,
and the calculated spectral function shows that the ``linear dispersion'' in PNSM is robust against disorder.
 It is also shown that the speed of the ballistic transport 
obeys the scaling theory \cite{Fradkin86critical1,Fradkin86critical2,Kobayashi14density,Fu17accurate,Roy18global} near the PNSM-metal transition point.

\iftitle\section{Lattice models} \label{sec:model}\fi
\iftitle
 \subsection{Dirac semimetal}
\else
 \textit{Model for Dirac semimetals.}
\fi
 For the numerical simulations, we consider the following two models of a simple cubic lattice describing disordered PNSMs.
 The DSM phase, where doubly degenerated 3D Dirac cones arise at the high-symmetry points in the Brillouin zone,
can be realized on the topological-trivial phase boundary of topological insulators.
 As a simple model for topological insulators,
we employ the Wilson-Dirac type tight-binding Hamiltonian \cite{Qi08topological,Liu10model,Ryu12disorder,Kobayashi13disordered,Kobayashi14density,Sbierski14z2,Kobayashi15dimensional,Yoshimura16comparative}, 
\begin{align}
 H\_D &=  \sum_{\bf r} \sum_{\mu=x,y,z} 
        \[ \ket{{\bf r}+{\bf e}_\mu} 
           \(  {iu \over 2}  \alpha_{\mu}
              -{m_2 \over 2} \beta 
           \)
           \bra{\bf r}  + \textrm{H.c.}
        \]  \nonumber \\
   & + \sum_{\bf r} \ket{\bf r} 
        \[(m_0 +3 m_2) \beta
           + V({\bf r}) 1_4
        \] \bra{\bf r},
 \label{eqn:H_DSM}
\end{align}
where ${\bf r}$ is the position of lattice sites and ${\bf e}_\mu$ 
($\mu = x,y,z$)
is the lattice vector in the $\mu$ direction.
 $u$ is the nearest neighbor transfer with spin-orbit interactions,
$m_0$ is referred to as ``mass'', and $m_2$ is the Wilson parameter.
 The length unit is set to the lattice constant.
 Dirac matrices $\alpha_\mu$ and $\beta$ are an anticommuting set of $4\times 4$ matrices 
satisfying $\alpha_\mu^2 = \beta^2 = 1_4$, with $1_4$ the identity matrix.
 We introduce an on-site random potential $V({\bf r})$, which is uniformly distributed in $[-{W\over 2},{W\over 2}]$.
 We take $m_2$ as the energy unit, and set $u/m_2 = 2$.
 We tune the parameters $(m_0/m_2, W/m_2)$ along the line of DSM phase 
starting from $(-2,0)$, 
where the Dirac points locate at $(k_x,k_y,k_z) = (\pi,0,0)$, $(0,\pi,0)$, and $(0,0,\pi)$.
 The DSM phase, which is the phase boundary between the weak and strong topological insulators,
is identified by the transfer matrix method \cite{Kobayashi13disordered}.
 Note that the slope of the Dirac cones (i.e.,~group velocity at the nodal point) is $u$ in the clean limit.

\iftitle\subsection{Weyl semimetal}
\else\textit{Model for Weyl semimetals. }\fi
 The time-reversal-broken type WSM phase, where pairs of 3D Dirac cones arise,
can be realized in 
the tight-binding Hamiltonian \cite{Yang11quantum,Imura11spin,Chen15disorder,Shapourian16phase,Yoshimura16comparative},
\begin{align}
 H\_W &=  \sum_{\bf r}
        \sum_{\mu=y,z}
         \[  \ket{{\bf r}+{\bf e}_\mu}
           \(  {iu \over 2}  \sigma_{\mu}
           \)
           \bra{\bf r}  + \textrm{H.c.}
         \] \nonumber \\
   &+  \sum_{\bf r}
        \sum_{\mu=x,y,z}
         \[ \ket{{\bf r}+{\bf e}_\mu}
           \(
              -{m_2 \over 2} \sigma_x
           \)
           \bra{\bf r}  + \textrm{H.c.}
         \]  \nonumber \\
   & + \sum_{\bf r} \ket{\bf r}
        \[(m_0 +3 m_2) \sigma_x
           + V({\bf r}) 1_2
        \] \bra{\bf r},
 \label{eqn:H_WSM}
\end{align}
where $\sigma_\mu$ are the Pauli matrices.
 The Weyl nodes are split in the $k_x$ direction with separation $2k_0$,
\begin{align}
 k_0 &=  \arccos{\( m_0/m_2+3 -\cos{k_y}-\cos{k_z} \)},
 \label{eqn:k0_WSM}
\end{align}
where $k_y,k_z=0$ or $\pi$.
 The parameters $u$ and $m_2$ are the same as in the DSM.
 We tune the parameters $(m_0/m_2, W/m_2)$ on the line starting from $(-1,0)$ where the Weyl points locate at $(k_0=\pm \pi/2,0,0)$ (details of the tuning are described later).
 The clean limit group velocity at the Weyl points is $m_2$ in the $x$ direction, and is $u$ in the $y$ and $z$ directions.

\iftitle
 \section{Time-evolution simulation}
\else
 \textit{Time-evolution simulation.}
\fi
 We study the dynamics of wave packets in PNSMs in the presence of randomness
by a direct time-evolution simulation on the lattice models.
 For simplicity, 
we focus on the transport in a specific ($x$) direction without loss of generality.
 Periodic boundary conditions are imposed in the directions transverse to the transport ($y$ and $z$),
and the system length is sufficiently large so that the boundaries in the $x$ direction can be neglected.
 That is, we consider a quasi-one-dimensional (Q1D) system with the cross-section of $L\times L$ sites.
 The 3D dynamics is obtained from the limit $L\to \infty$.

 The time-evolution operator $U(\Delta t)$ for the wave function $\psi(t)$,
\begin{align}
 \psi(t+\Delta t) = U(\Delta t) \psi(t),
 \label{eqn:U}
\end{align}
can be written for a time-independent Hamiltonian $H$,
\begin{align}
 U(\Delta t) =  \exp(-i H \Delta t),
 \label{eqn:eiHt}
\end{align}
where $\hbar=1$. 
 The exponential function is expanded by Chebyshev polynomial $T_n$ \cite{Tal-Ezer98an},
\begin{align}
 \exp(-i\tilde{H} \Delta \tilde{t}\,) \simeq  \sum_{n=0}^{M} (-i)^n C_{n}\, J_n(\Delta \tilde{t}\,)\, T_n(\tilde{H}),
 \label{eqn:Cheb}
\end{align}
where $M$ is an order of truncation \cite{note1}, $J_n$ is the Bessel function of the first kind, and $C_0=1$, $C_{n\ge 1}=2$.
 We have rescaled the Hamiltonian and time step so that the eigenvalues of $\tilde{H}$ are in the range $(-1,1)$,
\begin{align}
 \tilde{H} = { H \over \lambda\_{max} },\ 
 \Delta\tilde{t} = \lambda\_{max} \Delta t ,
 \label{eqn:resc}
\end{align}
where $\lambda\_{max}$ is set to be slightly larger than the possible maximum absolute value in eigenvalues of $H$.
 
 We focus on the time evolution of the Dirac/Weyl states near the nodal points.
 Thus we prepare the state corresponding to the eigenstate 
at a nodal point $\mathbf{k}_0$ in a clean system, $\exp(i\mathbf{k}_0 \cdot \mathbf{r})\chi$.
 Then we make an initial wave packet
by multiplying a Gaussian factor centered at $x_0$,
\begin{align}
 \psi(x,y,z,t=0) = \exp\[-{(x-x_0)^2\over{\xi^2}}\] \exp(i\mathbf{k}_0 \cdot \mathbf{r})\chi.
 \label{eqn:ini}
\end{align}
 We adopt the eigenstate of $\alpha_x$ for DSM and $\sigma_x$ for WSM as the spinor $\chi$ of the initial state,
so that the wave packet moves in the $x$ direction
according to the spin-momentum locking.
 We set the width of the wave packet $\xi=40$ for weak disorders and $\xi=10$ near the critical point, in order to see the wave-packet dynamics efficiently.

\begin{figure}[tbp]
 \centering
  \includegraphics[width=0.98\linewidth]{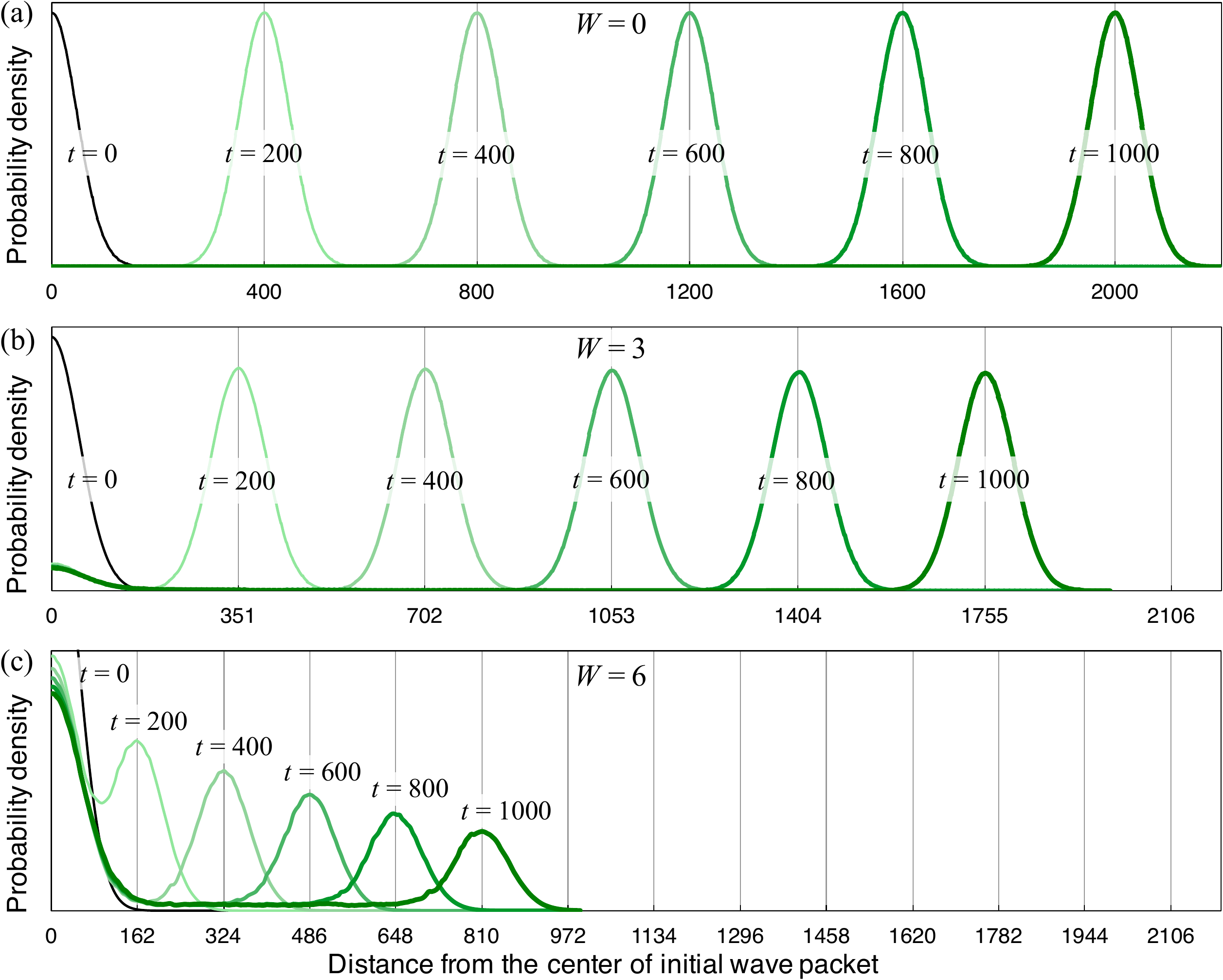}
\caption{
  (a) Time evolution of wave packets for DSM at (a) the clean limit $W=0$, (b) a weak disorder $W=3$, and (c) a strong disorder $W=6$.
 The vertical axis is 
$F(x,t)$, 
and the horizontal axis is distance from the center of the initial wave packet $x-x_0$.
 The size of the cross section $L=35$.
}
\label{fig:wavepacket}
\end{figure}

\iftitle\section{Ballistic transport in disordered DSM}\fi
\iftitle \subsection{Wavepacket dynamics}
\else \textit{Wavepacket dynamics. }\fi
 Figure \ref{fig:wavepacket} shows the time evolution of a wave packet
in disordered DSM.
 The plotted probability density $F(x,t)=\braket{\sum_{s}|\psi_s(x,y,z,t)|^2}_{y,z,\mathrm{dis}}$ is summed over all internal degrees of freedom, and averaged over the cross section at $x$
and over disorder realizations (e.g.,~$4000$ simulations for $W=6$).
 The wave packet travels linearly with time, that is, \textit{ballistic}, 
even in a strong disorder
(note that the DSM-metal transition occurs at $W=W\_c^{\mathrm{DSM}}\simeq 6.4$ \cite{Kobayashi14density}),
while the speed slows down with increasing disorder strength.
 The wave packets left
at the initial position in Figs.~\ref{fig:wavepacket}(b) and \ref{fig:wavepacket}(c) are the diffusive components in the initial wave packet.
 The height of the wave packet decays near $W\_c$, as the energy range of ballistic transport decreases and mixes with the diffusive states.
 The reduction of the ballistic range can be intuitively understood by spectral functions shown later.

\iftitle\subsection{Scaling behavior of speed}
\else\textit{Scaling behavior of speed. }\fi
 In a system with dispersion $E= v_\alpha |k|^\alpha$,
the number of states below $k$ behaves as 
\begin{align}
 \mathcal{N} \propto k^d \propto v_\alpha^{-d/\alpha} |E|^{d/\alpha},
 \label{eqn:NoS}
\end{align}
and the DOS is given by
\begin{align}
 \rho(E) \propto v_\alpha^{-d/\alpha} |E|^{d/\alpha-1}.
 \label{eqn:DoS}
\end{align}
 On the other hand, the scaling theory \cite{Kobayashi14density} predicts that near the critical disorder strength $W\_c$, that is, $\delta =\frac{|W\_c-W|}{W\_c} \simeq 0$, DOS obeys a scaling formula,
\begin{align}
 \rho(E) = \delta^{(d-z)\nu}f(|E| \delta^{-z\nu}),
 \label{eqn:rhoScal}
\end{align}
where $z$ is the dynamical exponent and $\nu$ is the critical exponent. 
 In the systems where $\alpha$ is well defined,
the scaling formula Eq.~\eqn{rhoScal} should have the same energy dependence as Eq.~\eqn{DoS},
\begin{align}
 \rho(E) \sim \delta^{-d(z/\alpha-1)\nu} |E|^{d/\alpha-1}.
 \label{eqn:rhoAlpha}
\end{align}
 Assuming the linear dispersion $\alpha=1$ in disordered PNSM phases, 
the scaling behavior of the speed of Dirac/Weyl electrons, $v=v_1$, 
is obtained by comparing Eq.~\eqn{DoS} with Eq.~\eqn{rhoAlpha},
\begin{align}
 v \sim \delta^{(z-1)\nu} \sim (W\_c-W)^{(z-1)\nu}.
 \label{eqn:vScal}
\end{align}

\begin{figure}[tbp]
 \centering
  \includegraphics[width=0.98\linewidth]{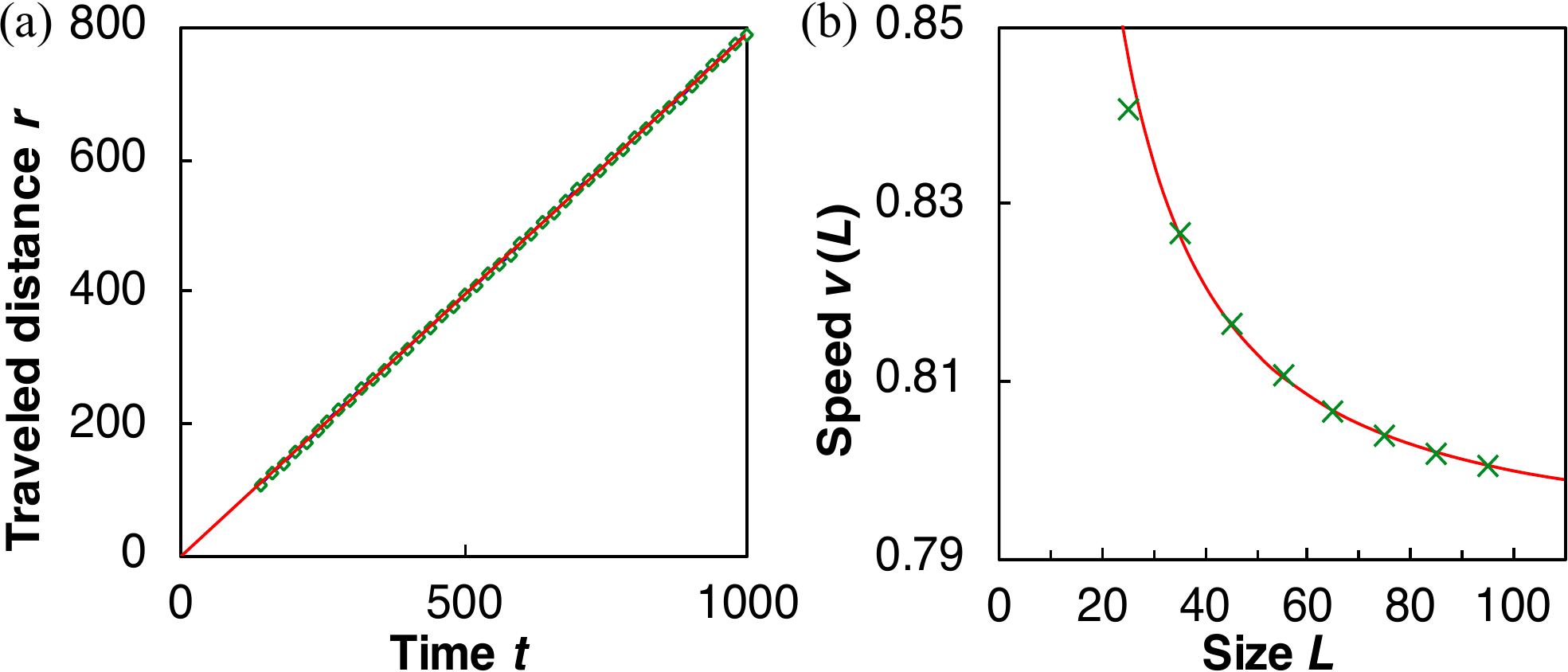}
 \vspace{-2mm}
\caption{
  (a) Traveled distance of the peak of wave packet $r$ as a function of time $t$ at $W=6$, $L=95$, averaged over 200 samples. 
  The solid line is a linear fitting.
  (b) Speed of ballistic mode $v(L)$ in Q1D DSM samples with $W=6$ as a function of cross-section size $L$.
  The solid line is a fitting curve Eq.~\eqn{vFit}.
}
\label{fig:vScalD}
\end{figure}

 We evaluated the travelled distance of wave-packet peak $r$
after smoothing the curves by taking averages,
and estimated the speed of the ballistic modes $v(L)=\frac{dr}{dt}$ for a specific size of cross section.
 In order to estimate the speed $v$ for the 3D limit, $L \to \infty$,
we estimated the speed for different sizes ($L=35,45,55,65,75,85,95$)
and fitted the data with the formula
\begin{align}
 v(L) = a_1 L^{-a_2} + v,
 \label{eqn:vFit}
\end{align}
where $a_1$ and $a_2>0$ are fitting parameters.
 Then the obtained speeds $v$ (see Fig.~\ref{fig:vScalD}) coincide with the scaling ansatz Eq.~\eqn{vScal} with $W\_c^{\mathrm{DSM}} = 6.4$, $\nu^{\mathrm{DSM}} = 0.8$, and $z=1.5$ \cite{Kobayashi14density}
near the critical point,
while they reproduce the slopes of the linear bands, $2$ and $1$ for DSM and WSM, respectively, in the clean limit.

\begin{figure}[tbp]
 \centering
  \includegraphics[width=0.98\linewidth]{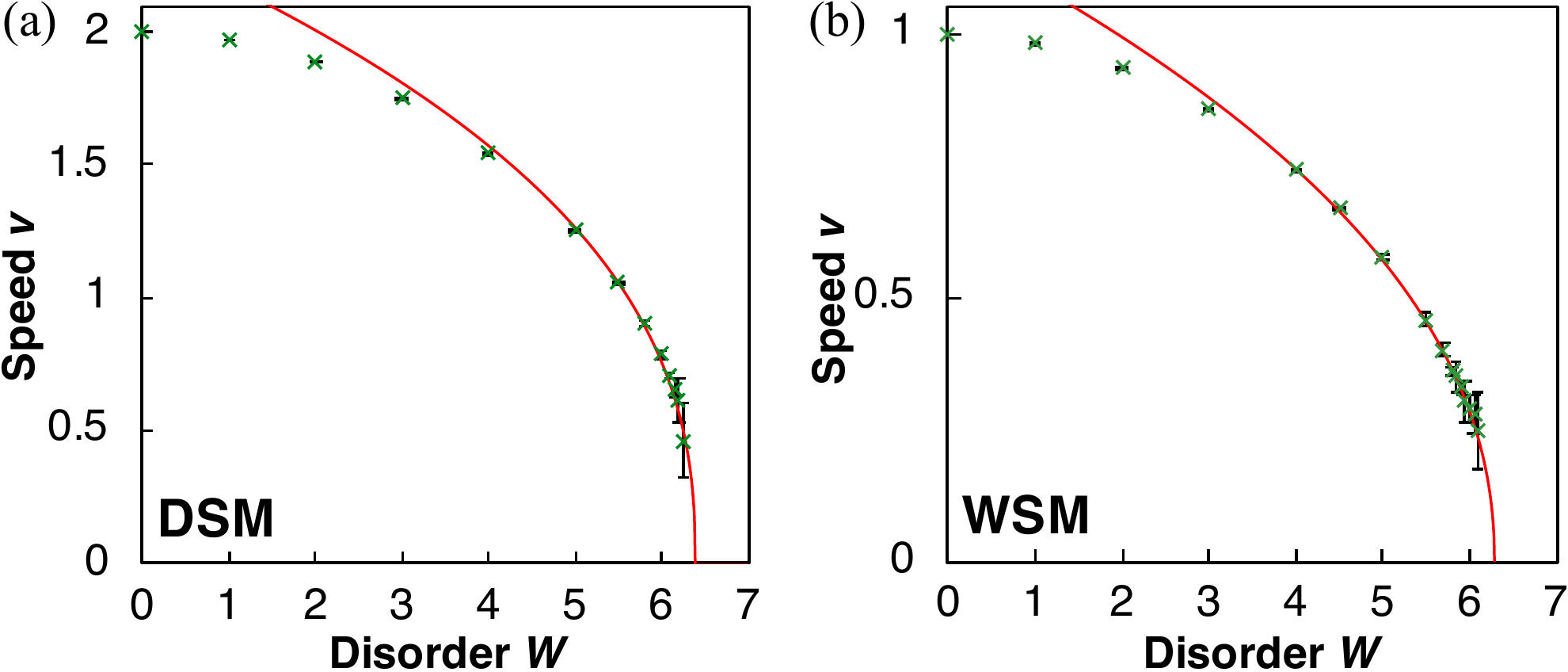}
 \vspace{-2mm}
\caption{
  Speed of ballistic mode in the 3D limit $v=v(L\to\infty)$ as a function of disorder strength $W$,
  in (a) DSM and (b) WSM.
  The solid line is a fitting curve Eq.~\eqn{vScal} with $W\_c = 6.4$ and $\nu=0.8$ for DSM, and $W\_c = 6.3$ and $\nu=0.9$ for WSM.
}
\label{fig:vScalW}
\end{figure}

 The ballistic transport is expected to be a common feature for PNSMs.
 We demonstrate the ballistic transport occurs also in the disordered WSMs where the point-nodes arise in pairs and the time-reversal symmetry is broken (class A \cite{Wigner51on,Dyson62statistical,Altland97nonstandard}),
while the DSM preserves the time-reversal symmetry (class AII).
 The behavior of the wave packets and of their speed is qualitatively the same as in the case of DSM.
 We estimated the speed of wave packets dynamics in WSMs in the same way (see Fig.~\ref{fig:vScalW}).
 Assuming $z=1.5$ \cite{Goswami11quantum,Kobayashi14density,Roy14diffusive,Roy16erratum,Pixley16uncovering}, the speed can be fitted by Eq.~\eqn{vScal} with 
$W\_c^{\mathrm{WSM}}=6.3$, $\nu^{\mathrm{WSM}}=0.9$, which is consistent with Refs.~\onlinecite{Liu16effect,Bera16dirty}.

\begin{figure}[tbp]
 \centering
  \includegraphics[width=1\linewidth]{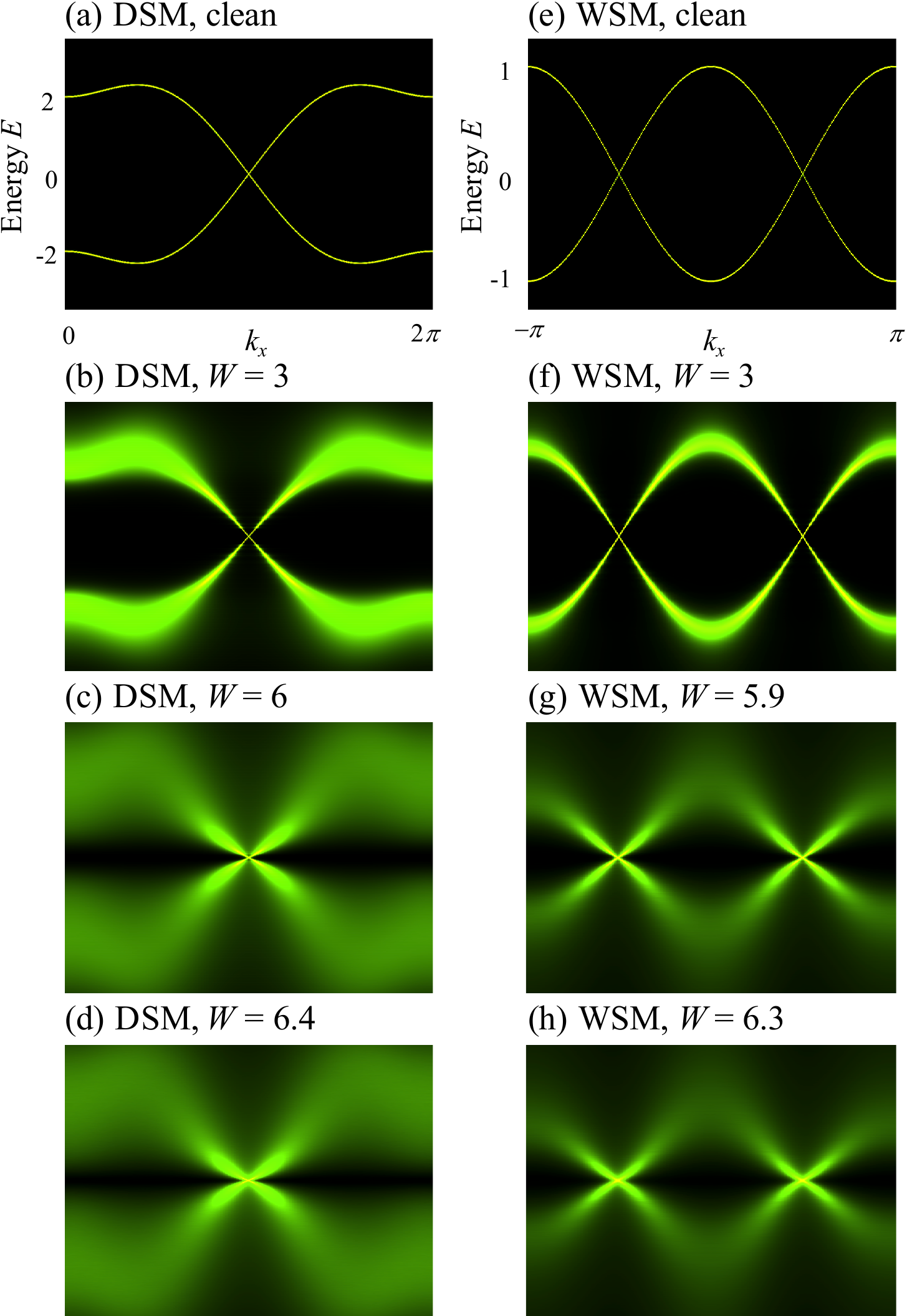}
 \vspace{-3mm}
\caption{
  Spectral feature $\rho_{(k_x,0,0)}(E)$ of disordered DSM (a)-(d) and WSM (e)-(h) with different disorder strength: 
  (a)(e) clean, (b)(f) weak, (c)(g) strong, and (d)(h) critical.
  The horizontal axis is $k_x$ and the vertical axis is $E$.
  The system size is $360\times 50 \times 50$ sites for DSM and $360\times 100 \times 100$ sites for WSM.
  Averaged over up to $6$ samples.
}
\label{fig:NARPES}
\end{figure}

\iftitle\subsection{Robustness of nodal points}
\else \textit{Robustness of nodal points. }\fi
 The presence of a ballistic mode under disorder implies that the states near the nodal points are robust against disorder.
 Here, we visualize this numerically.
 In disordered systems, we cannot define the dispersion and therefore the $k$-space position of nodal points, in principle.
 However, a spectral feature of ``linear dispersion'' and ``nodal point'' in real (say, disordered) materials
is observed by angle-resolved photoemission spectroscopy (ARPES) measurements.
 The spectral function $\rho_{\bm{k}}(E)$, which is equivalent to the imaginary part of the retarded Green's function, $-\frac{1}{\pi} \mathrm{Im} \mathrm{Tr}_s G\^R(E,\bm{k},s)$, can be understood as the local density of states in momentum space,
\begin{align}
 \rho_{\bm{k}}(E)
  = \frac{1}{NS} \sum_{s=1}^{S} \braket{\bm{k},s|\delta(E-H)|\bm{k},s},
 \label{eqn:spect}
\end{align}
where $N$ is the number of lattice sites and $S$ the internal degrees of freedom (four in DSM and two in WSM).
 We can obtain the spectral function by using the kernel polynomial method \cite{Silver96kernel,Weisse06KPM}, which enables large-scale calculations using the Chebyshev polynomial expansion,
\begin{align}
 &\rho_{\bm{k}}(\tilde{E}) \simeq \frac{1}{\pi\sqrt{1-\tilde{E}}} \sum_{n=0}^{M}C_n g_n \mu_n T_n(\tilde{E}),\\
  &g_n\! =\! \tfrac{1}{M+2}\!\[\! (M\!-\!n\!+\!2)\cos\!\tfrac{\pi n}{M+2} + \sin\!\tfrac{\pi n}{M+2}\cot\!\tfrac{\pi}{M+2} \]\!,\\
  &\mu_n = \frac{1}{NS} \sum_{s=1}^{S} \braket{\bm{k},s|T_n(\tilde{H})|\bm{k},s},
 \label{eqn:KPM}
\end{align}
where $\tilde{E}=E/\lambda\_{max}$.
 Here we focus on the bulk spectral function
by taking $\braket{\mathbf{r}|\bm{k},s} = \psi_{\bm{k},s}(x,y,z)$ as
\begin{align}
 \psi_{\bm{k},s}(x,y,z) = e^{ik_x x} e^{ik_y y} e^{ik_z z} \chi_s,
 \label{eqn:planeWave}
\end{align}
where $\chi_s$ is one of the orthonormal bases.
 Figure~\ref{fig:NARPES} shows the density plot of the spectral function Eq.~\eqn{spect} at $k_y=k_z=0$.
 In the clean system $W=0$, the dispersion is reproduced [Figs.~\ref{fig:NARPES}(a) and \ref{fig:NARPES}(b)].
 With increasing disorder, the ``band'' gets blurred.
 However, we can still find a clear ``nodal point'' and ``linear dispersion''
below the critical disorder strength of the DSM-metal transition $W\_c^{\mathrm{DSM}}\simeq 6.4$
or the WSM-metal transition $W\_c^{\mathrm{WSM}}\simeq 6.3$.
 The slope of the ``linear dispersion'' decreases with increasing disorder;
the evolution of the slope is consistent with the self-consistent Born approximation \cite{Ominato14quantum,Ominato15quantum,Liu16effect,Luo18unconventional}
and coincides with that of the speed of wave-packet dynamics (shown in Fig.~\ref{fig:vScalW}).
 In addition, the ``gap'' of cosine bands, i.e., the energy range where only the ``linear dispersion'' arises, is narrowed with increasing disorder;
the cosine bands are broadened by the disorder and the ``gap'' closes in a diffusive metallic phase ($W > W\_c$).
 We note that the surface spectral function, which is accessible by the ARPES measurements,
can be obtained in the same way and gives the same dependence on the disorder.

 We have also utilized the spectral function to fix the effective mass for the disordered WSMs.
 The spectrum in WSMs (as in DSMs) shows the linear band structures and nodal points
up to the critical point,
and enables us to estimate the position of Weyl nodes.
 We have tuned the mass $m_0$ so that the Weyl nodes locate at $(\pm \pi/2,0,0)$ with a sufficient precision.

\iftitle\subsection{Critical Superdiffusion}
\else \textit{Critical superdiffusion. }\fi
 Before concluding, we mention some interesting properties 
that signal the PNSM-metal phase transition.
 The first point is the vanishing velocity of the ballistic modes (see Fig.~\ref{fig:vScalW}).
 The second point is the breakdown of the linear dispersion [see Figs.~\ref{fig:NARPES}(d) and \ref{fig:NARPES}(h)].
 The third point is superdiffusion.
 The scaling formula for the mean displacement at $E=0$ is \cite{Kobayashi14density}
\begin{align}
 \braket{||\bm{r}||} \sim \delta^{-\nu} f_r(t \delta^{z\nu}).
 \label{eqn:rScal}
\end{align}
 At the critical point $\delta=0$,
$\delta$ dependence should be cancelled to avoid singularity, 
\begin{align}
 \braket{||\bm{r}||} \sim t^{1/z}.
 \label{eqn:rScalcricical}
\end{align}
 Since $1/z\simeq 0.67>1/2$,
a superdiffusion is expected at the critical point.
 Although it was technically difficult to see the critical superdiffusion directly (since the wave-packet speed vanishes at the critical point),
a sub-ballistic transport has been obtained in the vicinity of the critical point.
 It can be also seen in Fig.~\ref{fig:NARPES}
as a superlinear ``dispersion'' $2>\alpha>1$ around the critical point.

\iftitle\section{Conclusions} \label{sec:conclusion}
\else \textit{Conclusion. }\fi
 We have studied the wave-packet dynamics of disordered DSM/WSMs
and found that the PNSMs show a ballistic transport,
reflecting the robustness of PNSMs against disorder.
 Similar ballistic transport is also observed in topological insulators on a percolative lattice \cite{Mano19application}, 
and is a universal behavior.
 It is to be noted that since the wave packet studied here is not an energy eigenstate,
the ballistic transport feature is not restricted to $E = 0$.
 This is due to the fact that the life time of ballistic particles is long even if it is away from $E=0$, 
and the renormalization of the speed is the same for $E=0$ and $|E|>0$, 
which can be seen in the robust ``linear dispersion'' (spectral function) of disordered PNSMs.
 The finite life time may be reflected by the amplitude of the ballistic wave packets, which diminishes as they travel.
 We have confirmed that the speed of the ballistic mode slows down and vanishes toward the PNSM-metal transition point.
 The behavior of the speed is consistent with the prediction of the scaling theory.
 This means that the speed of the ballistic mode can be an order parameter characterizing the PNSM-metal transition.
 Indeed, we have estimated the critical point and the critical exponent for the WSM-metal transition.
 The approach we introduced will be applied to other types of semimetals and transitions, such as the surface of topological insulators or the corner of higher-order topological insulators, and semimetal-insulator transitions or semimetal-semimetal transitions.
 It might be also interesting to study the ballistic motion in one-dimensional semimetal-metal transitions \cite{Garttner15disorder}.

\iftitle\begin{acknowledgements}\fi
 We thank Kentaro Nomura and Yuya Ominato for useful discussions.
 This work was supported by 
KAKENHI (Grants 
No.~JP16J01981, 
No.~JP19K14607, and 
No.~JP19H00658) 
from the Japan Society for the Promotion of Science.
\iftitle\end{acknowledgements}\fi

\bibliography{Ballistic}

\begin{thebibliography}{61}
\bibitem [{not()}]{note1}%
  \BibitemOpen
  \href@noop {} {}\bibinfo {note} {The Bessel function $J_n(p)$ decays
  monotonically for $n\gg p$, with asymptotic behavior $J_n(p)\sim
  \frac{1}{\sqrt{2\pi n}}(\frac{ep}{2n})^n$. We set $M$ ($\gg \Delta
  \tilde{t}$) such that $J_M(\Delta \tilde{t})<10^{-15}$.}\BibitemShut {Stop}%
\end{thebibliography}

\end{document}